# Local Codes with Cooperative Repair in Distributed Storage System

Jing Wang[1], Zhiyuan Yan[2], Hongmei Xie[3]

1. School of Information Engineering, Chang'an University, Xi'an 710064, China;
2. Department of Electrical and Computer Engineering, Lehigh University, Bethlehem, PA 18015, USA;
3. Seagate Technology, Fremont, CA 94538 USA;

**Abstract:** Recently, the research on local repair codes is mainly confined to repair the failed nodes within each repair group. But if the extreme cases occur that the entire repair group has failed, the local code stored in the failed group need to be recovered as a whole. In this paper, local codes with cooperative repair, in which the local codes are constructed based on minimum storage regeneration (MSR) codes, is proposed to achieve repairing the failed groups. Specifically, the proposed local codes with cooperative repair construct a kind of mutual interleaving structure among the parity symbols, that the parity symbols of each local code, named as distributed local parity, can be generated by the parity symbols of the MSR codes in its two adjacent local codes. Taking advantage of the structure given, the failed local groups can be repaired cooperatively by their adjacent local groups with lower repair locality, and meanwhile the minimum distance of local codes with cooperative repair is derived. Theoretical analysis and simulation experiments show that, compared with codes with local regeneration (such as MSR-local codes and MBR-local codes), the proposed local codes with cooperative repair have benefits in bandwidth overhead and repair locality for the case of local groups failure.

**Keywords:** distributed storage system, local codes with cooperative repair, minimum storage regeneration (MSR) codes, storage overhead, repair bandwidth overhead, repair locality

## I. Introduction

In large distributed storage systems (DSSs), node failure is inevitable along with data loss. To retain high availability of DSSs, Dimakis et al. introduce regenerating codes based on network coding [1], which can reduce maintenance bandwidth use more compared with a hybrid of replication and erasure codes. Ernvall give the construction of regenerating codes between the minimum storage regenerating (MSR) points and the minimum bandwidth regenerating (MBR) points [2, 3].

Apart from storage overhead and repair bandwidth overhead, repair locality, the number of nodes contacted by the replacement node in the repair process [4-6], is an important metric to measure the cost of data repair. The repair locality relates to the disk input/output (I/O) overhead, which has been the main performance bottleneck during repairing failed nodes.

To reduce the disk I/O overhead, Papailiopoulos et al. proposed simple regenerating codes [7] that employ simple XORs over the MDS coded packets to perform exact repair, and further present an optimal locally repairable codes (LRCs) based on simple combinations of RS codes that can achieve arbitrarily high data rates and better repair locality [8]. Based on maximum rank distance (MRD) Gabidulin codes, Rawat and Silberstein construct the optimal LRCs with two-layer encoding structure [9, 10], and derive the upper bound on the amount of data stored in DSSs. However, in view of the properties of Gabidulin codes [11], the complexity of the optimal LRCs will increase exponentially with the number of nodes in DSSs.

Moreover, Kamath et al. focus on the construction of optimal codes with local regeneration [12], which combine the advantages of both LRCs and regenerating codes, aiming to simplify node repair and decrease encoding complexity. The constructed optimal codes with local regeneration, such as MSR-local codes and MBR-local codes, contain multiple local codes and one global parities part, in which the local codes are either MSR codes or MBR codes. Considering the extreme case that one entire repair group has failed, the corresponding local code in the failed group inevitably needs to be recovered as a whole, and the corresponding local group is named as the failed group. MSR-local codes and MBR-local codes proposed by Kamath et al. [12] have the ability to repair the failed local codes, even though their initial purpose was to simplify node repair. Nevertheless, from the construction of MSR-local codes and MBR-local codes, MSR-local codes and MBR-local codes can repair only one failed local code, incapable of repairing two or more failed local codes at the same time. Furthermore, it is necessary for MSR-local codes and MBR-local codes to collect all the remaining local codes and the global parities for repairing the only failed local code, which will add the repair complexity.

RAID 4 consists of block-level striping with a dedicated parity disk, whereas RAID 5 with distributed parity [13]. Unlike in RAID 4, RAID 5's parity information is distributed among the drives, which evens out the stress of a dedicated parity disk among all RAID members. Inspired by the distributed parity of RAID 5, the global parities of MSR-local codes may also be broken down into multiple distributed local parities to avoid the limitations of repair failed local codes above as MSR-local codes. In this paper, we present an explicit construction of local codes with cooperative repair (LCCR), which only contains local codes, and no global parities. Compared with MSR-local codes or MBR-local codes, each local code of LCCR includes MSR code part and distributed local parity part. Concretely, MSR code part in LCCR can be expressed in the form of systematic code including information symbols and parity symbols. Further, the distributed local parity part of each local code in LCCR can be generated by the parity symbols of the MSR codes in its two adjacent local codes, which can be regarded as a kind of mutual interleaving structure among the parity symbols. On the basis of the structure of the proposed LCCR, the failed local groups can be recovered cooperatively by their adjacent groups with lower repair locality, and the minimum distance of LCCR is derived. Theoretical and MATLAB data analyses show that, the proposed LCCR have performance improvement in bandwidth overhead and repair locality for the cases of local group failure. But LCCR has more storage overhead than MSR-local codes, since each local code of LCCR has the distributed local parity part.

The remainder of this paper is organized as follows. In Section II, we discuss relevant background and related work. Explicit construction of LCCR is presented in Section III, followed by deducing the minimum distance of LCCR. Section IV provides the performance analyses, including repair of a single failed node, repair of one failed local group and repair of multiple failed local groups. And Section V concludes the paper.

## II. Background and Related Work

### A. MSR codes and MBR codes

In a DSS, a message file of $M$ symbols over $GF(q)$ is dispersed across $n$ active storage nodes, each of which has a storage capacity of $\Gamma$ code symbols. Suppose a newcomer receives the same amount of information from each of the existing nodes, any $k$ storage nodes can recover the original file. When a storage node failed, a replacing node collects $s$ symbols each from any $d$ surviving nodes, and hence the total repair bandwidth is $x = ds$.

Clearly, since we can reconstruct the entire original file from any $k$ storage nodes, any node can be recovered in the DSS. In this way, a total bandwidth of $k\Gamma \geq M$ bits is required to repair anyone failed node with capacity of $\Gamma$ symbols. According to the results in [1], by connecting to $d$ surviving nodes, less bandwidth of $x = ds$ will be required to recover the failed node. If the size of the file is fixed to be $M$, there exists an optimal tradeoff between the storage overhead per node $\Gamma$ and the repair bandwidth $x$. The codes which attain the optimal tradeoff above is called regenerating codes. And the codes that attain the extremal point of minimum storage overhead is named as minimum-storage regenerating (MSR) codes, with

$$(\Gamma_{MSR}, x_{MSR}) = \left( \frac{M}{k}, \frac{Md}{k(d-k+1)} \right).$$

Similarly, the codes that attain the extremal point of minimum repair bandwidth is named as minimum-bandwidth regenerating (MBR) codes, with

$$(\Gamma_{MBR}, x_{MBR}) = \left( \frac{2Md}{2kd - k^2 + k}, \frac{2Md}{2kd - k^2 + k} \right).$$

The repair process above can be either functional-repair or exact-repair [2, 3]. For functional-repair, the nodes may change over time, i.e., if a node $v_i^{old}$ is lost and in the repair process, we get a new node $v_i^{new}$ instead, then we may have $v_i^{old} \neq v_i^{new}$. Contrarily, exact-repair means that the nodes does not vary over time, that is the new node $v_i^{new}$ is always the same as the old one $v_i^{old}$. Exact-repair can obviate additional communication overheads during the repair process, and also avoid retuning of the reconstruction and repair algorithm. In this paper, exact-repair is considered to maintain the local codes in systematic form during the repair operation.

### B. MSR-local codes and MBR-local codes

By combining the advantages of LRCs as well as regenerating codes, Kamath et al. proposed codes with local regeneration [12], which are codes with locality over a vector alphabet. The constituent local codes themselves are regenerating codes, such as MSR codes or MBR codes. Concretely, the construction of MSR-local codes is shown in Fig.1.

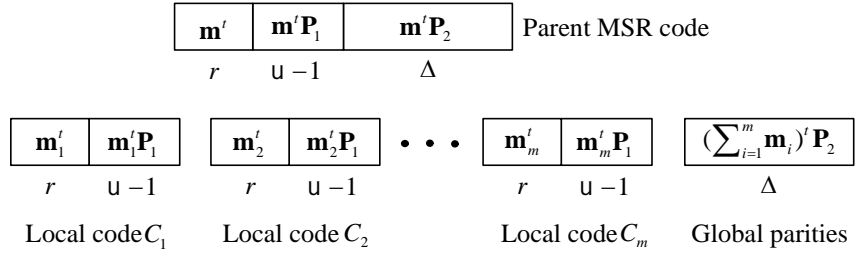

Fig. 1 The construction of MSR-local codes

It is worth noting that the parent MSR code $\mathbf{m}^t, \mathbf{m}^t\mathbf{P}_1, \mathbf{m}^t\mathbf{P}_2$ is $((n_L+\Delta, r, d), (\mathsf{r}, \mathsf{s}))$ MSR code with $n_L = r+\mathsf{u}-1$ and $d \leq r+\mathsf{u}-2$, which generator matrix is $\mathbf{G}_0 = [\mathbf{G}_L | \mathbf{Q}_\Delta]$. Puncturing the parent MSR code to the first $n_L$ symbols, we can obtain an $((n_L, r, d), (\mathsf{r}, \mathsf{s}))$ MSR code $\mathbf{m}^t, \mathbf{m}^t\mathbf{P}_1$. The MSR-local codes $\mathbf{C}_{\text{MSR-local}}$ can be achieved by taking MSR code $\mathbf{m}_i^t, \mathbf{m}_i^t\mathbf{P}_1$ as local code $C_i$ ($1 \leq i \leq m$), and $(\sum_{i=1}^m \mathbf{m}_i)^t \mathbf{P}_2$ as global parities. The corresponding generator matrix $\mathbf{G}_{\text{MSR-local}}$ is given by

$$\mathbf{G}_{\text{MSR-local}} = \begin{bmatrix} \mathbf{G}_L & & & \mathbf{Q}_\Delta \\ & \ddots & & \vdots \\ & & \mathbf{G}_L & \mathbf{Q}_\Delta \end{bmatrix}.$$

Similarly, the MBR-local codes can be realized by the same method as MSR-local codes. The MBR-local codes $\mathbf{C}_{\text{MBR-local}}$ proposed in [12] also compose of $m$ support-disjoint MBR codes and $\Delta$ global parities. Specially, each MBR code in MBR-local codes is a repair-by-transfer (RBT) $((n_L, r, d), (\mathsf{r}, \mathsf{s}), K_L)$ MBR code [14]. Thus the desired MBR-local codes $\mathbf{C}_{\text{MBR-local}}$ will have length $n = mn_L + \Delta$, and scalar rank $K = mK_L$.

## III. Construction of Local Codes with Cooperative Repair

For MSR-local codes and MBR-local codes, it is required to collect all the remaining local codes and the global parities to repair one failed local code, with high repair complexity. Inspired by the distributed parity of RAID 5, the global parities of MSR-local codes may also be broken down into multiple distributed local parities to avoid the limitations of repairing the failed local codes. Thus, we attempt to construct local codes with cooperative repair (LCCR), in which each local code includes MSR code part and the distributed local parity part.

**Construction 1:** Let $C_i'$ ($1 \leq i \leq m$) be an $((n_L, r, d), (\mathsf{r}, \mathsf{s}), K_L = r\mathsf{r})$ exact-repair MSR code with generator matrix $\mathbf{G}_i = [\mathbf{I} | \mathbf{P}_i]$, where $n_L = r+\mathsf{u}-1$ such that $d \leq r+\mathsf{u}-2$. Based on the MSR code above, an $[n, K, d_{\min}, \mathsf{r}]$ LCCR $\mathbf{C}$ is constructed with $n = m(n_L + \Delta)$ and $K = mK_L$, in which the parameters $\mathsf{u}, \Delta$ are chosen such that $\mathsf{u} \geq \Delta$. LCCR $\mathbf{C}$ has $m$ local codes $\{C_1, C_2, C_3, \cdots, C_m\}$, each of which includes MSR code part and distributed local parity part. Generator matrix $\mathbf{G}$ of the desired LCCR $\mathbf{C}$ is

$$\mathbf{G} = \begin{bmatrix} \mathbf{G}_1 & \mathbf{0} & \mathbf{0} & \mathbf{P}_1 & \mathbf{0} & \mathbf{0} & \cdots & \mathbf{0} & \mathbf{P}_1 \\ \mathbf{0} & \mathbf{P}_2 & \mathbf{G}_2 & \mathbf{0} & \mathbf{0} & \mathbf{P}_2 & \cdots & \mathbf{0} & \mathbf{0} \\ \mathbf{0} & \mathbf{0} & \mathbf{0} & \mathbf{P}_3 & \mathbf{G}_3 & \mathbf{0} & \cdots & \mathbf{0} & \mathbf{0} \\ \mathbf{0} & \mathbf{0} & \mathbf{0} & \mathbf{0} & \mathbf{0} & \mathbf{P}_4 & \cdots & \mathbf{0} & \mathbf{0} \\ \vdots & \vdots & \vdots & \vdots & \vdots & \vdots & \ddots & \vdots & \vdots \\ \mathbf{0} & \mathbf{0} & \mathbf{0} & \mathbf{0} & \mathbf{0} & \mathbf{0} & \cdots & \mathbf{0} & \mathbf{P}_{m-1} \\ \mathbf{0} & \mathbf{P}_m & \mathbf{0} & \mathbf{0} & \mathbf{0} & \mathbf{0} & \cdots & \mathbf{G}_m & \mathbf{0} \end{bmatrix}$$

In particular, $C_i$ ($1 < i < m$) can be generated by the submatrix

$$\begin{bmatrix} \mathbf{0} & \mathbf{0} \\ \vdots & \vdots \\ \mathbf{0} & \mathbf{P}_{i-1} \\ \mathbf{G}_i & \mathbf{0} \\ \mathbf{0} & \mathbf{P}_{i+1} \\ \vdots & \vdots \\ \mathbf{0} & \mathbf{0} \end{bmatrix}$$

And the local code $C_1$ (or $C_m$) can be generated by the submatrix composed of the first two columns of generator matrix $\mathbf{G}$ (or the last two columns). According to the matrix construction above, the distributed local parity symbols of each local code can be created by the parity symbols of the MSR codes in its two adjacent local codes, which can be regarded as a kind of mutual interleaving structure among the parity symbols.

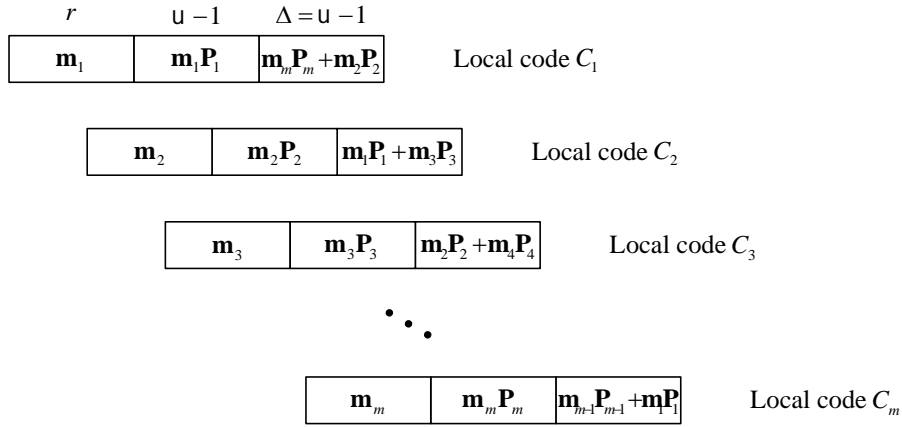

Fig. 2 The construction of LCCR $\mathbf{C}$

Fig. 2 illustrates the corresponding code construction of LCCR $\mathbf{C}$. From Fig. 2, local code $C_i$ includes MSR code part $\mathbf{m}_i$, $\mathbf{m}_i\mathbf{P}_i$ and distributed local parity part $\mathbf{m}_{i-1}\mathbf{P}_{i-1} + \mathbf{m}_{i+1}\mathbf{P}_{i+1}$, where $\mathbf{m}_{i-1}\mathbf{P}_{i-1}$ and $\mathbf{m}_{i+1}\mathbf{P}_{i+1}$ are the parity symbols of the MSR code parts in its two adjacent local codes. It need to point out that local code $C_i$ has two adjacent local codes $C_{(i-2)_m+1}$ and $C_{(i)_m+1}$, where $(i)_m$ is an integer in $\{0,1, 2, \cdots, m-1\}$ such that $(i)_m \equiv i \mod m$.

The theorem below identifies the parameters of LCCR so constructed and proves the construction to yield LCCR with minimum distance $d_{\min}$.

**Theorem 1:** Consider LCCR $\mathbf{C}$ constructed in Construction 1 in which the parameters $u$ and $\Delta$ are chosen such that $\Delta = u - 1$. Then LCCR $\mathbf{C}$ with $(r, u)$ information locality has length $n = m(n_L + \Delta)$ and scalar dimension $K = mK_L$. The minimum distance of LCCR $\mathbf{C}$ is given by

$$d_{\min} = n - \frac{K}{r} + u - 2\left(\frac{K}{r\Gamma} - 1\right)(u-1) = u + 2\Delta$$

*Proof:* We will prove the minimum distance $d_{\min}$ of LCCR $\mathbf{C}$ is $u + 2\Delta$ as the expression above provided. Since $n = m(r + u - 1 + \Delta)$, $K = mr\Gamma$ and $\Delta = u - 1$, we have that

$$\begin{aligned}d_{\min} &= n - \frac{K}{r} + u - 2\left(\frac{K}{r\Gamma} - 1\right)(u-1) \\ &= n - mr + u - 2(m-1)(u-1) = u + 2\Delta\end{aligned}$$

Towards LCCR $\mathbf{C}$ in Fig. 2, it suffices to show that any non-zero codeword $\mathbf{C}$ has Hamming weight $wt(\mathbf{C}) \geq u + 2\Delta$. First of all, note that if $\mathbf{C}$ has non-zero components belonging to one or more local codes, then $wt(\mathbf{C}) \geq u + 2\Delta$, since all MSR code parts in local codes have minimum distance $u$ and the distributed local parity parts have minimum distance $\Delta$. Next, consider the complementary case where the non-zero components of $\mathbf{C}$ are restricted to one MSR code part and two relevant local parity parts of local codes, such as the MSR code part $\mathbf{m}_1, \mathbf{m}_1\mathbf{P}_1$ and two parity parts $\mathbf{m}_1\mathbf{P}_1$ in local code $C_2$ and $\mathbf{m}_1\mathbf{P}_1$ in $C_m$. And when the all-zero code symbols are deleted from each codeword, then the resultant punctured codeword $\mathbf{C}_{punc}$ lies in the space of the resulting matrix

$$\mathbf{G}_{punc} = \begin{bmatrix} \mathbf{G}_1 & \mathbf{P}_1 & \mathbf{P}_1 \\ 0 & 0 & 0 \\ 0 & 0 & 0 \\ 0 & 0 & 0 \\ \vdots & \vdots & \vdots \\ 0 & 0 & 0 \\ 0 & 0 & 0 \end{bmatrix}$$

which is reduced from the generator matrix $\mathbf{G}$. As the information vector remains unchanged and still is $[\mathbf{m}_1 \ \mathbf{m}_2 \ \mathbf{m}_3 \ \cdots \ \mathbf{m}_m]$, the punctured codeword $\mathbf{C}_{punc}$ is $[\mathbf{m}_1\mathbf{G}_1 \ \mathbf{m}_1\mathbf{P}_1 \ \mathbf{m}_1\mathbf{P}_1]$, and its Hamming weight $wt(\mathbf{C}_{punc}) = u + 2\Delta$. By the analysis above, it can be followed that the minimum distance $d_{\min}$ of LCCR is $u + 2\Delta$.

## IV. Performance Comparison

In this section, we choose a representative set of codes, including LCCR, MSR-local codes and MBR-local codes, to analyze their performances in storage overhead, repair bandwidth overhead and repair locality. Taking into account that the bandwidth overhead and the repair locality will change with the concrete failures of local groups or nodes, the cases of a single node failure, one local group failure, and multiple local groups failure are discussed respectively.

*A. Storage overhead*

In this paper, we adopt the definition of storage overhead $n\mathsf{r}/K$ given in [12], which is defined to be the ratio of the total number of storage units required to store the encoded symbols over the number of message symbols contained in the original file. We can obtain the storage overhead of LCCR, MSR-local codes and MBR-local codes separately by

$$S_{\text{LCCR}} = \frac{n\mathsf{r}}{K} = \frac{n_L + \Delta}{r}$$

$$S_{\text{MSR-local}} = \frac{n\mathsf{r}}{K} = \frac{n_L + \Delta/m}{r}$$

$$S_{\text{MBR-local}} = \frac{n\mathsf{r}}{K} = \frac{2(n_L + \Delta/m)(r + \mathsf{u} - 2)}{r(r + 2\mathsf{u} - 3)}$$

Considering MSR-local codes and MBR-local codes have the similar code structure with code length $n = mn_L + \Delta$ and minimum distance $d_{\min} = \mathsf{u} + \Delta$, the parameter $m$ can be achieved with the same value for MSR-local codes and MBR-local codes as $n$ and $d_{\min}$ fixed. Similarly, the parameters $r$, $\mathsf{u}$, $\Delta$ in MSR-local codes and MBR-local codes also have the same values. Thus, we can compare the storage overhead of MSR-local codes with that of MBR-local codes directly. For MBR-local codes, its storage overhead

$$S_{\text{MBR-local}} = \frac{2(r + \mathsf{u} - 2)}{(r + 2\mathsf{u} - 3)} \cdot S_{\text{MSR-local}} = \mathsf{y} \cdot S_{\text{MSR-local}}.$$

Since in MBR-local codes, scalar dimension

$$K = m \cdot \left( r\mathsf{r} - \binom{r}{2} \right) \geq 1,$$

and thus $r \geq 2$. Meanwhile with $\mathsf{u} \geq 3$, it can be deduced that

$$\mathsf{y} = \frac{2(r + \mathsf{u} - 2)}{(r + 2\mathsf{u} - 3)} > 1.$$

The storage overhead of MBR-local codes is lower bounded by that of MSR-local codes, and the factor $\mathsf{y}$ reflects the extra storage overhead of MBR-local codes compared with MSR-local codes.

As fixing the values of code length $n$ and minimum distance $d_{\min}$, the values of the parameters $m$, $r$, $\mathsf{u}$, $\Delta$ obtained in LCCR will differ from that of MSR-local codes and MBR-local codes, since its code length $n = m(n_L + \Delta)$ and its minimum distance $d_{\min} = \mathsf{u} + 2\Delta$. Thus, we cannot compare the storage overhead of LCCR with that of MSR-local codes, or that of MBR-local codes directly according the formulae defined by $n\mathsf{r}/K$. In next subsection, the comparison about the storage overhead of LCCR, MSR-local codes, or MBR-local codes will be plotted by MATLAB data analysis as $n = 120$ and $d_{\min} = 16$.

*B. Repair of a single failed node*

To maintain the integrity of codes, even though the failed nodes are in the parity part, either the global parities of MSR-local codes or MBR-local codes, or the distributed local parity of LCCR, the failed nodes are all considered to be repaired in this paper. We only discuss the case of one single node failure, not including the case of multiple nodes failure, which can be regarded as a generalization of the case of one node failure.

Although the proposed LCCR intends to repair the failed local codes as a whole, LCCR, which is constructed based on MSR codes, has the ability to repair the failed nodes in local groups. When the failed nodes of one local group are in the MSR code part, we can adopt the same method as MSR codes to repair. If there are several failed nodes in the distributed local parity part of one local code, we can repair the failed nodes by collecting the parity parts of the MSR codes in its two adjacent local codes. Concretely for LCCR, take as an example that there exists a single failed node in the distributed local parity $\mathbf{m}_{i-1}\mathbf{p}_{i-1} + \mathbf{m}_{i+1}\mathbf{p}_{i+1}$ of local code $C_i$, shown as in Fig. 3. To repair the single failed node, local group $i$ should collect symbols $\mathbf{m}_{i-1}\mathbf{p}_{i-1}$ from local group $i-1$, and symbols $\mathbf{m}_{i+1}\mathbf{p}_{i+1}$ from local group $i+1$. Furthermore by simple xoring $\mathbf{m}_{i-1}\mathbf{p}_{i-1}$ and $\mathbf{m}_{i+1}\mathbf{p}_{i+1}$, the distributed local parity part $\mathbf{m}_{i-1}\mathbf{p}_{i-1} + \mathbf{m}_{i+1}\mathbf{p}_{i+1}$ can be achieved, and meanwhile the single failed node in distributed local parity $\mathbf{m}_{i-1}\mathbf{p}_{i-1} + \mathbf{m}_{i+1}\mathbf{p}_{i+1}$ can be repaired.

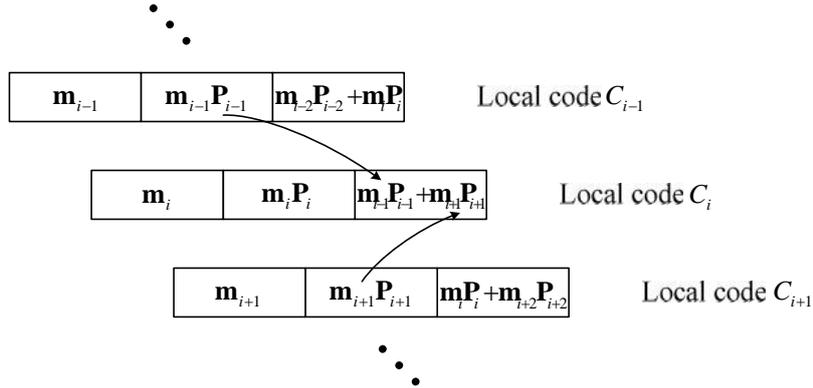

Fig. 3 Repair of a single failed node located in the distributed local parity of LCCR

Furthermore, we discuss the performance of repair locality for LCCR, MSR-local codes and MBR-local codes in the case of a single node failure. Due to the construction of LCCR that the failed nodes may be located in the MSR code part of local codes, or in the distributed local parity part, we should consider the two cases together.

**Lemma 1:** The locality to repair one single failed node located in the MSR code part of LCCR is $n_L - 1$.

**Lemma 2:** The locality to repair one single failed node located in the distributed local parity part of LCCR is $2(u-1)$.

**Theorem 2:** The locality of LCCR to repair one single failed node is

$$\frac{(n_L - 1) \cdot n_L + 2(u-1) \cdot \Delta}{n_L + \Delta}$$

Since the local codes of MSR-local codes is MSR codes, the repair locality of MSR-local codes is the same as that of MSR codes when repairing one single failed node in local codes, which equals $n_L - 1 = r + u - 2$. Meanwhile, if the single failed node is located in the global parities part, the global parities of MSR-local codes should be repaired as a whole, which needs $mr$ nodes in total. The repair locality of MSR-local codes is

$$\frac{(n_L - 1) \cdot m n_L + mr \cdot \Delta}{m n_L + \Delta} = \frac{(n_L - 1) \cdot n_L + r \cdot \Delta}{n_L + \Delta/m}$$

Adopting the same method, we can obtain the repair locality of MBR-local codes which equals that of MSR-local codes, as the formula describes above.

Using MATLAB data analysis, we can compare the locality of LCCR with that of MSR-local codes, or that of MBR-local codes. Concretely, we choose common length $n = 120$ and common minimum distance $d_{\min} = 16$. For comparison conveniently, we set the parameter $m \geq 3$ since there is at least $m \geq 3$ local codes in LCCR. Fig. 4 illustrates the repair locality versus storage overhead for LCCR, MSR-local codes and MBR-local codes. From Fig. 4, MBR-local codes has the same repair locality as that of MSR-local codes as they have the same values for parameter $m$, $r$, $u$, $\Delta$, which is consistent with the theoretical analysis. Through MATLAB data analysis, LCCR has the smallest repair locality under the same storage consumption.

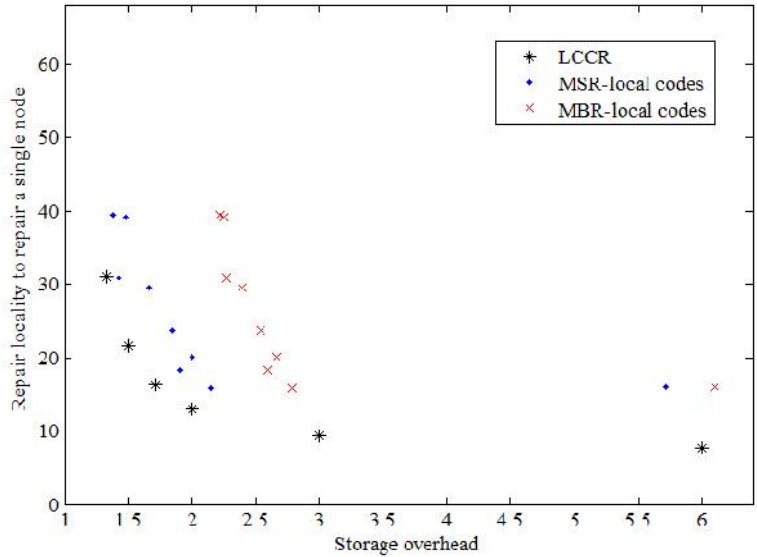

Fig. 4 Repair locality versus storage overhead for one single node failure

Next, we analyze the bandwidth overhead to repair one single failed node. The normalized bandwidth overhead can be calculated by $n\check{S}/K$ [12], where $\check{S}$ denotes the average repair bandwidth for repairing a single failed node.

**Theorem 3:** The bandwidth overhead of LCCR to repair one single failed node is

$$B_{\text{LCCR}} = \frac{n \cdot \check{S}_{\text{LCCR}}}{K} = \frac{n_L^2 (n_L - 1)}{r^2 (u - 1)} + \frac{2(u - 1)^2}{r}.$$

*Proof:* According to the repair process of a single failed node in LCCR, the failed node may be located in the MSR code part of local codes, or in the distributed local parity part. We can calculate $\check{S}_{LCCR}$ by

$$\check{S}_{LCCR} = \frac{\check{S}_1 \cdot n_L + \check{S}_2 \cdot \Delta}{n_L + \Delta}$$

where $\check{S}_1 = \frac{n_L \Gamma}{r} \cdot \frac{n_L - 1}{n_L - r}$ denotes the average repair bandwidth for repairing a single failed node located in the MSR code part, and $\check{S}_2 = 2(u-1)\Gamma$ denotes the repair bandwidth required for repairing one failed node located in the distributed local parity part. It is worth noting that the repair bandwidth needed is always $2(u-1)\Gamma$ whether there are one or more failed nodes in distributed local parity, since the distributed local parity part should be repaired as a whole at any time. According to the analysis above, the bandwidth overhead of LCCR is

$$B_{LCCR} = \frac{n \cdot \check{S}_{LCCR}}{K} = \frac{n_L^2(n_L - 1)}{r^2(u-1)} + \frac{2(u-1)^2}{r}.$$

For MSR-local codes and MBR-local codes, all the information symbols of the $m$ local codes should be collected together to repair one single failed node located in the global parities, which will further increase the repair bandwidth consumption. The repair bandwidth overhead of MSR-local codes and MBR-local codes can also be calculated by $n\check{S}/K$ [12].

**Lemma 3:** For MSR-local codes, the average repair bandwidth for repairing a single failed node located in local codes is $\check{S}_1' = \frac{n_L \Gamma}{r} \cdot \frac{n_L - 1}{n_L - r}$, and the average repair bandwidth for repairing a failed node located in global parities part $\check{S}_2' = mK = mr\Gamma$. The average repair bandwidth

$$\check{S}_{MSR\text{-}local} = \frac{\check{S}_1' \cdot m \cdot n_L + \check{S}_2' \cdot \Delta}{m \cdot n_L + \Delta},$$

The bandwidth overhead of MSR-local codes to repair one single failed node is

$$B_{MSR\text{-}local} = \frac{n \cdot \check{S}_{MSR\text{-}local}}{K} = \frac{n_L^2(n_L - 1)}{r^2(u-1)} + \Delta.$$

**Lemma 4:** In MBR-local codes, each MBR code adopted is a repair-by-transfer (RBT) $((n_L, r, d), (\Gamma, s), K_L)$ MBR code [14]. The average repair bandwidth for repairing a single failed node located in local codes is $\check{S}_1'' = n_L - 1$, and the average repair bandwidth for repairing a failed node located in global parities part $\check{S}_2'' = mK_L = m \cdot \left(r\Gamma - \binom{r}{2}\right)$. The bandwidth overhead of MBR-local codes for repairing one single failed node is

$$B_{MBR\text{-}local} = \frac{n \cdot \check{S}_{MBR\text{-}local}}{K} = \frac{2n_L(n_L - 1)}{r(r + 2u - 3)} + \Delta.$$

A summary of repair locality and bandwidth overhead for LCCR, MSR-local codes and MBR-local codes is presented in Table 1, as one single node failed. For the same reason that MSR-local codes and MBR-local codes have the similar code structure, we can compare their bandwidth overhead conveniently, obtaining that $B_{MBR\text{-}local} < B_{MSR\text{-}local}$. Similarly, we adopt MATLAB data analysis to compare the bandwidth overhead of LCCR with that of MSR-local codes, or that of MBR-local codes. As $n = 120$, $d_{min} = 16$ and $m \geq 3$, the performance of LCCR, MSR-local codes and MBR-local codes in repair bandwidth versus storage overhead is shown in Fig. 5.

Table 1 Summary of repair locality and bandwidth overhead for LCCR, MSR-local codes and MBR-local codes for one single node failure

| Construction | Repair locality | Bandwidth overhead |
|---|---|---|
| LCCR | $\dfrac{(n_L-1)\cdot n_L + 2(\mathrm{u}-1)\cdot \Delta}{n_L + \Delta}$ | $\dfrac{n_L^2(n_L-1)}{r^2(\mathrm{u}-1)} + \dfrac{2(\mathrm{u}-1)^2}{r}$ |
| MSR-local Codes | $\dfrac{(n_L-1)\cdot n_L + r\cdot \Delta}{n_L + \Delta/m}$ | $\dfrac{n_L^2(n_L-1)}{r^2(\mathrm{u}-1)} + \Delta$ |
| MBR-local Codes | $\dfrac{(n_L-1)\cdot n_L + r\cdot \Delta}{n_L + \Delta/m}$ | $\dfrac{2n_L(n_L-1)}{r(r+2\mathrm{u}-3)} + \Delta$ |

Since MSR-local codes and MBR-local codes achieve the same values for parameter $m$, $r$, $\mathrm{u}$, $\Delta$ while fixing $n=120$ and $d_{\min}=16$, we will obtain the same code number for MSR-local codes and MBR-local codes. Specially in Fig. 5, MSR-local codes and MBR-local codes have similar distribution, furthermore which are mutually corresponding. It is obvious that MBR-local codes and MSR-local codes will be separately located in the corresponding position of their distribution region of the plot, when they have the same value for parameter $m$, $r$, $\mathrm{u}$, $\Delta$. The storage overhead of MBR-local codes is obviously larger than that of MSR-local codes, and meanwhile its repair bandwidth is smaller than that of MSR-local codes, in accordance with the theoretical analysis. From Fig. 5, the storage overhead of LCCR is smaller than that of most MBR-local codes, and is flat with that of MSR-local codes on the whole. At the same time, its bandwidth consumption is smaller than that of most MSR-local codes of the plot.

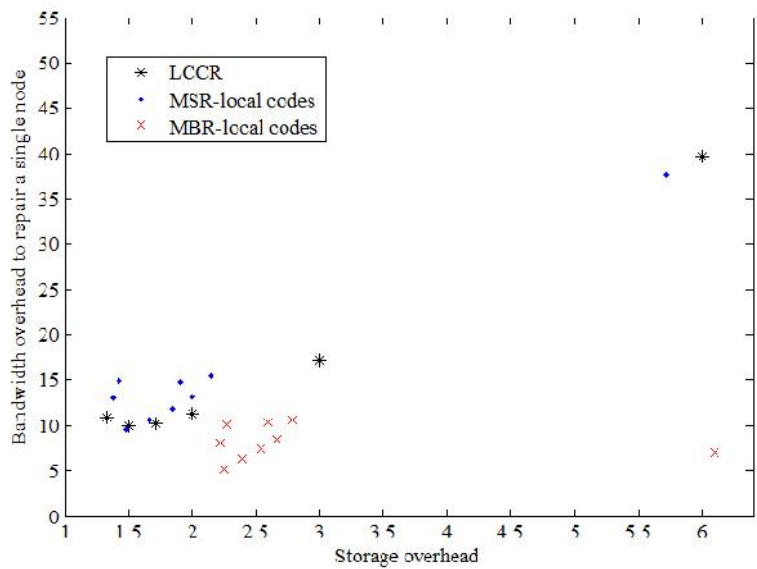

Fig. 5 Repair bandwidth versus storage overhead for one single node failure

*C. Repair of one failed local group*

Since the minimum distance $d_{\min\_local}$ of one local code in LCCR is $u$, the failed nodes cannot be repaired by other available nodes of the same local group as the number of failed nodes in one local code exceeds $u-1$. Thus some local groups, in which there exist more than $u-1$ failed nodes, are considered as failed local groups. For LCCR, one failed local group can be repaired cooperatively by its three adjacent local groups. Fig. 6 illustrates the concrete repair procedure for one failed local group corresponding to local code $C_i$ ($1 \leq i \leq m$). It can be seen that local code $C_i$ can be recovered by local codes $C_{i-2}$, $C_{i-1}$ and $C_{i+1}$. Collecting data $\mathbf{m}_{i-2}\mathbf{P}_{i-2}$ of local code $C_{i-2}$ in local group $i-2$, local group $i-1$ can recover data $\mathbf{m}_i\mathbf{P}_i$ by xoring $\mathbf{m}_{i-2}\mathbf{P}_{i-2}$ and $\mathbf{m}_{i-2}\mathbf{P}_{i-2}+\mathbf{m}_i\mathbf{P}_i$, and send the recovered data $\mathbf{m}_i\mathbf{P}_i$ to local group $i$. In view that the MSR code part of local code $C_i$ includes $\mathbf{m}_i$ and $\mathbf{m}_i\mathbf{P}_i$, the information data $\mathbf{m}_i$ can be easily obtained by using the generator matrix $\mathbf{G}_i=[\mathbf{I}\,|\,\mathbf{P}_i]$ of the MSR code $C'_i$ as $u-1 \geq r$. Furthermore, local group $i$ collect data $\mathbf{m}_{i-1}\mathbf{P}_{i-1}$ and $\mathbf{m}_{i+1}\mathbf{P}_{i+1}$ from local group $i-1$ and $i+1$, and gain the distributed local parity part $\mathbf{m}_{i-1}\mathbf{P}_{i-1}+\mathbf{m}_{i+1}\mathbf{P}_{i+1}$ of local code $C_i$ by simple xoring. Until now, we achieve the MSR code part and the distributed local parity part of local code $C_i$, and complete the repair of local group $i$. In the same way, we can also use local code $C_{i-1}$, $C_{i+1}$ and $C_{i+2}$ to recover local code $C_i$, and the recovery process is similar to that of Fig. 6.

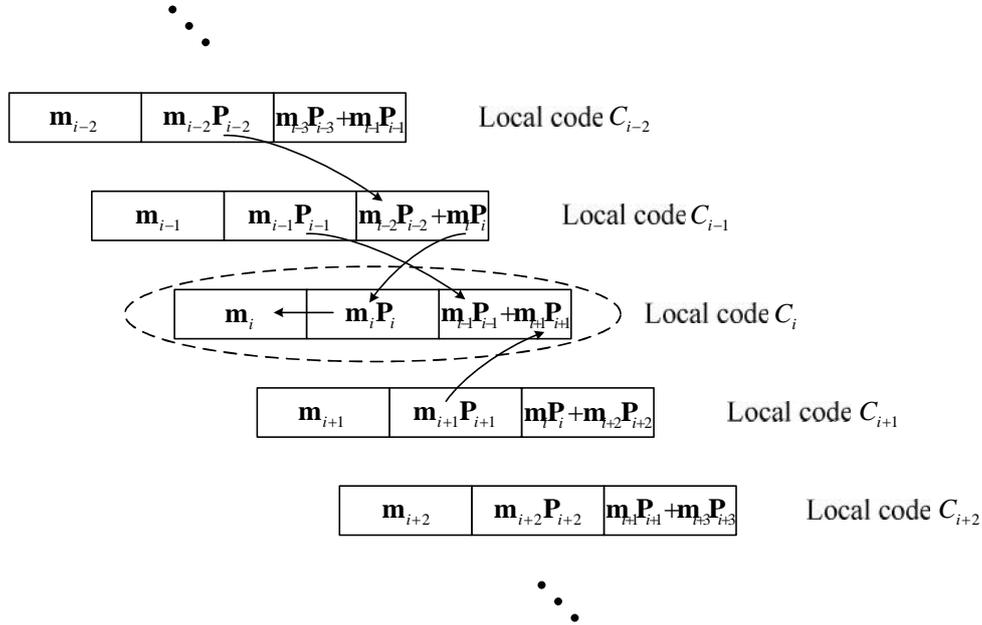

Fig. 6 Repair of one failed local group

**Definition 1 (locality for repairing failed local groups):** Similar to repairing failed nodes, the number of local groups that participate in the repair of failed local groups is defined as the repair locality for repairing failed local groups. Especially for MSR-local codes and MBR-local codes, the global parities part should be regarded as a whole during repairing.

From Fig. 6, any one failed local group of LCCR can be recovered by three local groups, and in this case the repair locality of LCCR to repair one local group is 3. For MSR-local codes and MBR-local codes, it is essential to collect all the surviving local codes and the global parities to repair the only one failed local group. Since MSR-local codes have $m$ local codes and one global parities part, the repair locality of MSR-local codes is $m$. As the same reason, the repair locality of MBR-local codes is also $m$. Consequently, MSR-local codes and MBR-local codes have the same repair locality, shown as in Fig. 7. From Fig. 7, when the storage overhead is smaller, MSR-local codes and MBR-local codes have the same repair localities of 3 as that of LCCR. As the storage overhead increases, the repair localities of MSR-local codes and MBR-local codes increase gradually until much larger than that of LCCR.

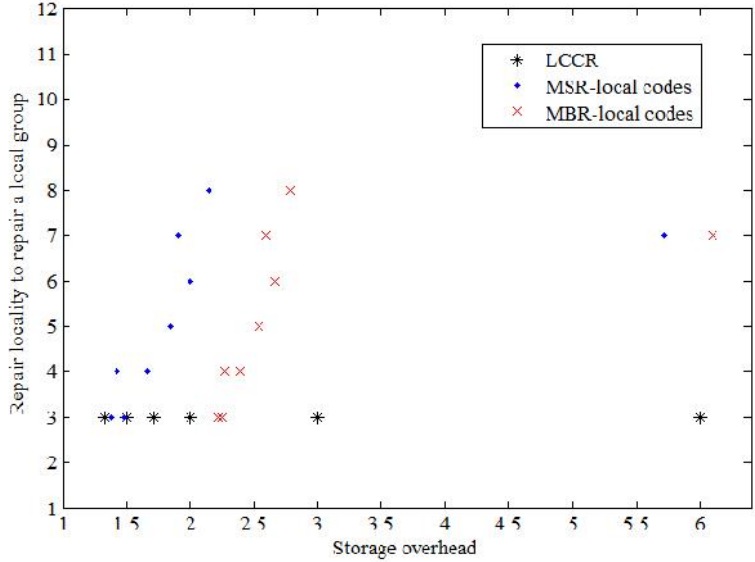

Fig. 7 Repair locality versus storage overhead for one local group failure

Similarly, the bandwidth overhead can be calculated by $(m\check{S}' + \check{S}'_\Delta)/K$, where $\check{S}'$ denotes the average repair bandwidth for repairing a single failed local group, and $\check{S}'_\Delta$ the repair bandwidth required for repairing the global parities (specially, $\check{S}'_\Delta = 0$ for LCCR).

**Lemma 5:** In LCCR, $\check{S}'_{\text{LCCR}} = 5(u-1)\Gamma$. Then the bandwidth overhead of LCCR is calculated by

$$B'_{\text{LCCR}} = \frac{m \cdot \check{S}'_{\text{LCCR}}}{K} = \frac{5(u-1)}{r}.$$

**Lemma 6:** In MSR-local codes, $\check{S}'_{\text{MSR-local}} = mr\Gamma + \Delta\Gamma$ and $\check{S}'_\Delta = mr\Gamma$. The bandwidth overhead of MSR-local codes takes on the form

$$B'_{\text{MSR-local}} = \frac{m\check{S}'_{\text{MSR-local}} + \check{S}'_\Delta}{K} = \frac{\Delta}{r} + m + 1.$$

**Lemma 7:** The average bandwidth overhead of MBR-local codes to repair one local group can be achieved as

$$B'_{\text{MBR-local}} = \frac{m\check{S}'_{\text{MBR-local}} + \check{S}'_{\Delta}}{K} = \frac{2\Delta(r+u-2)}{r(r+2u-3)} + m + 1,$$

with $\check{S}'_{\text{MBR-local}} = m\left(r\Gamma - \binom{r}{2}\right) + \Delta\Gamma$ and $\check{S}'_{\Delta} = m\left(r\Gamma - \binom{r}{2}\right)$.

Table 2 Summary of repair locality and bandwidth overhead for LCCR, MSR-local codes and MBR-local codes for one local group failure

| Construction | Repair locality | Bandwidth overhead |
|---|---|---|
| LCCR | 3 | $\frac{5(u-1)}{r}$ |
| MSR-local Codes | $m$ | $\frac{\Delta}{r} + m + 1$ |
| MBR-local Codes | $m$ | $\frac{2\Delta(r+u-2)}{r(r+2u-3)} + m + 1$ |

A summary of repair locality and bandwidth overhead for one local group failure is presented in Table 2. The bandwidth overhead of MBR-local codes is a little bigger than that of MSR-local codes as repairing one local group. MBR-local codes have special local codes, RBT MBR codes. In RBT MBR codes, any two nodes in a fully connected graph have a common code symbol, and the repair process can be accomplished by mere transfer of data and without any arithmetic operation. In the same way, we adopt MATLAB data analysis to compare the bandwidth overhead of LCCR with that of MSR-local codes, or MBR-local codes, at the case of one single local group failure.

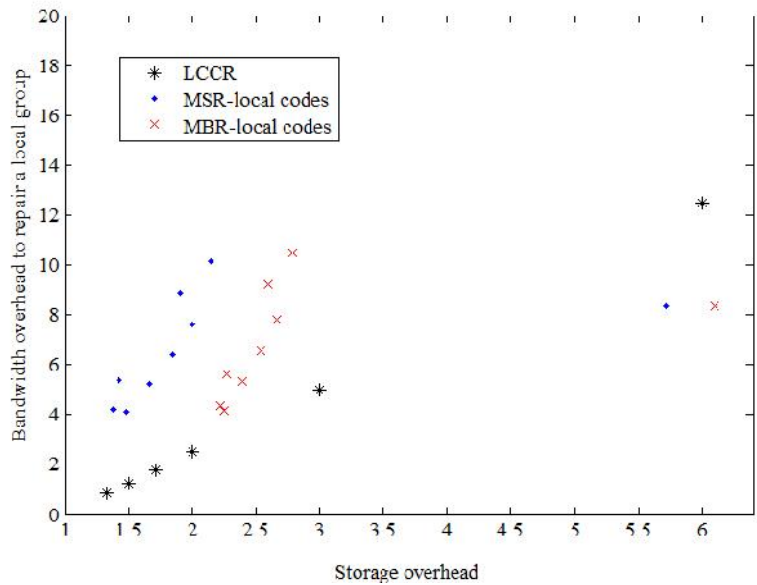

Fig. 8 Repair bandwidth versus storage overhead for one local group failure

Similarly choosing $n=120$, $d_{\min}=16$ and $m\geq 3$, we compare repair bandwidth versus storage overhead of LCCR, MSR-local codes and MBR-local codes for the case of one local group failure in Fig. 8. As storage overhead smaller, the bandwidth overhead of LCCR is much smaller than that of MSR-local codes. However, when LCCR consumes the largest storage overhead, its bandwidth overhead exceeds the bandwidth consumption of MSR-local codes and MBR-local codes. From Fig. 8, the bandwidth overhead of MBR-local codes is a little smaller than that of MSR-local codes, and mostly flat with that of MSR-local codes on the whole. The results obtained from Fig.8 are in accordance with the theoretical analysis above.

In Fig. 8, as the storage overhead of LCCR increases up to 6, the performance of LCCR in bandwidth overhead is worse than that of MSR-local codes and MBR-local codes. From the definition of storage overhead that is the ratio of the total number of encoded symbols over the number of message symbols contained in the original file, then the reciprocal of storage overhead can be regarded as the code rate $R$. It can be obtained that when the code rate of LCCR is much smaller, LCCR has not advantages on bandwidth overhead. Thus, LCCR with higher code rate should be chosen in distributed storage systems to achieve better performance in bandwidth consumption.

*D. Repair of multiple failed local groups*

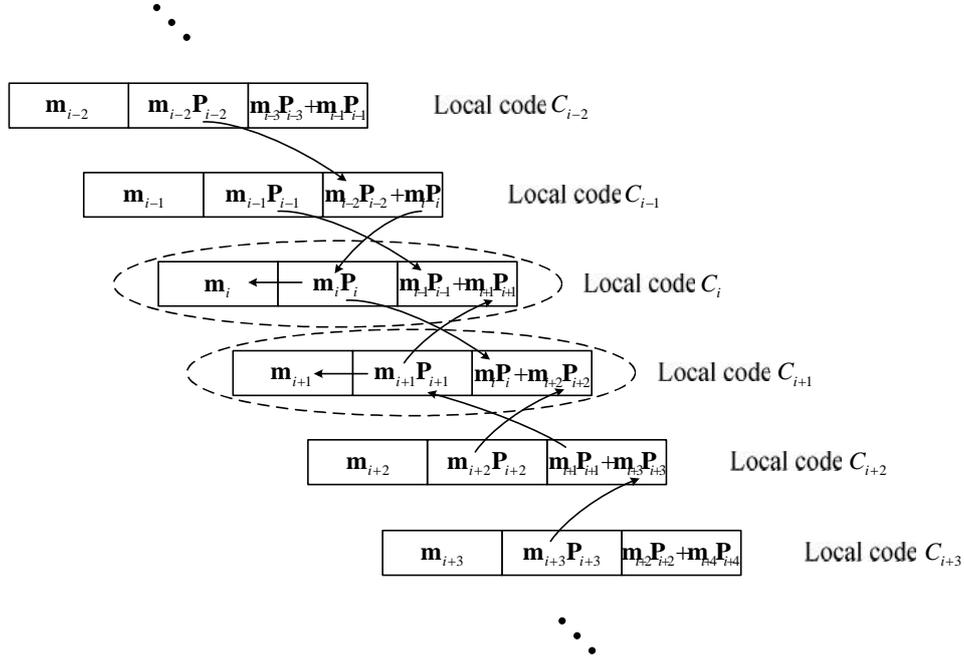

Fig. 9 Repair of two adjacent failed local groups

In this subsection, we discuss repair of multiple failed local groups, although the probability that multiple groups failed at the same time is very small. From the construction of LCCR in Fig. 2, several failed local groups of LCCR can also be recovered simultaneously. First, consider two adjacent local groups $i$ and $i+1$ failed, and the corresponding local code $C_i$ and $C_{i+1}$ need to be recovered, as shown in Fig. 9. From Fig. 9, we can get that the failed local group $i$ and $i+1$ can be recovered by four local groups $i-2$, $i-1$, $i+2$ and $i+3$. It can be deduced that when two adjacent local groups failed in LCCR, the repair locality of LCCR for repairing one failed local group is 2 on average. Furthermore, repairing two non-adjacent failed groups can be regarded as a generalization of repairing a single failed group, where the repair locality is 3 for repairing each failed local group.

Next, we consider the case that three local groups failed. Three failed local groups cannot be repaired only when the three local groups are adjacent to each other. For other cases that three local groups failed, up to 9 and at least 5 local groups are required to repair the failed local groups, which repair locality is up to 3 and at least 5/3 on average. Since the probability that four local groups failed simultaneously is very little, we can omit this case in this paper.

**Theorem 4 (Upper bound):** In LCCR, the number of failed local groups, which can be repaired cooperatively by other available local groups, is upper bounded by $u + 2\Delta - 1$.

*Proof:* Since LCCR can be regarded as one linear code with code length $n = m(r + u + \Delta - 1)$ and minimum distance $d_{\min} = u + 2\Delta$, at most $d_{\min} - 1 = u + 2\Delta - 1$ erasures in LCCR can be recovered simultaneously. If we distribute the $u + 2\Delta - 1$ erasures into $u + 2\Delta - 1$ local groups, then each local group will have one erasure. Then the number of repairable local groups is upper bounded by $u + 2\Delta - 1$.

Nevertheless, according to the structure of MSR-local codes and MBR-local codes proposed in [12], MSR-local codes and MBR-local codes have the ability to repair one failed local group, incapable of repairing two or more failed local groups at the same time. Moreover, for MSR-local codes and MBR-local codes, it is essential to collect all the remaining local codes and the global parities for repairing the only one failed local group, which will increase the repair complexity. Consequently, there exist limitations for MSR-local codes and MBR-local codes to recover the failed local groups.

## V. Conclusions

In this paper, we mainly investigate the case that in some local groups there exist more than $u - 1$ failed nodes, and the corresponding local codes stored need to be recovered as a whole. Concretely, an explicit construction of LCCR based on MSR codes is proposed to repair the failed local groups with lower bandwidth overhead and lower repair locality, in which the distributed local parity symbols of each local code can be generated by the parity symbols of the MSR codes in its two adjacent local codes, regarded as a kind of mutual interleaving structure among the parity symbols. Based on LCCR, the failed local groups can be repaired cooperatively by their adjacent local groups, and the minimum distance of LCCR is derived.

The performances of LCCR, MSR-local codes and MBR-local codes in storage overhead, repair bandwidth overhead and repair locality are discussed respectively. Theoretical and MATLAB data analyses show that, compared with MSR-local codes and MBR-local codes, LCCR has benefits in repair bandwidth overhead and repair locality for the case of one single local group failure.

# References

[1] A. G. Dimakis, P. B. Godfrey, Yunnan Wu, et al., "Network coding for distributed storage systems", IEEE Transactions on Information Theory, 2010, 56(9): 4539-4551.

[2] T. Ernvall, "Exact-regenerating codes between MBR and MSR points", 2013 IEEE Information Theory Workshop (ITW), Sevilla, Sept. 9-13, 2013: 1-5.

[3] T. Ernvall, "Codes between MBR and MSR points with exact repair property", IEEE Transactions on Information Theory, 2014, 60(11): 6993-7005.

[4] F. Oggier, A. Datta, "Self-repairing homomorphic codes for distributed storage systems", IEEE INFOCOM 2011, Shanghai, China, April 2011, pp. 1215-1223.

[5] O. Khan, R. Burns, J. Plank, et al., "In search of I/O-optimal recovery from disk failures", Proceedings of the 3rd USENIX conference on Hot topics in storage and file systems (HotStorage'11), Portland, OR, Jun. 2011.

[6] P. Gopalan, C. Huang, H. Simitic, et al., "On the locality of codeword symbols", IEEE Trans. on Information Theory, 2012, 58(11): 6925-6934.

[7] D. S. Papailiopoulos, Jianqiang Luo, A. G. Dimakis, et al., "Simple regenerating codes: network coding for cloud storage", The 31st Annual IEEE International Conference on Computer Communications: Mini-Conference.

[8] D. S. Papailiopoulos, A. G. Dimakis, "Locally repairable codes", 2012 IEEE International Symposium on Information Theory Proceedings (ISIT), pp. 2771-2775, Cambridge, MA, July 1-6, 2012.

[9] A. S. Rawat, N. Silberstein, O. O. Koyluoglu, S. Vishwanath, "Optimal locally repairable codes with local minimum storage regeneration via rank-metric codes", 2013 Information Theory and Applications Workshop (ITA), pp. 1-8, San Diego, CA, Feb. 10-15, 2013.

[10] N. Silberstein, A. S. Rawat, O. O. Koyluoglu, S. Vishwanath, "Optimal locally repairable codes via rank-metric codes", 2013 IEEE International Symposium on Information Theory (ISIT), pp.1819-1823, Istanbul, July 7-12, 2013.

[11] Antonia Wachter-Zeh, Valentin Afanassiev, Vladimir Sidorenko, "Fast decoding of Gabidulin codes", Designs, Codes and Cryptography, 2013, 66(1): 57-73.

[12] G. M. Kamath, N. Prakash, V. Lalitha, et al., "Codes with local regeneration and erasure correction", IEEE Transactions on Information Theory, 2014, 60(8): 4637-4659.

[13] Spencer W Ng, Mattson R L, "Maintaining good performance in disk arrays during failure via uniform parity group distribution", Proceedings of the First International Symposium on High-Performance Distributed Computing (HPDC-1), Syracuse, NY, 1992: 260-269.

[14] Shah N B, Rashmi K V, Kumar P V, Ramchandran K, "Distributed storage codes with repair-by-transfer and nonachievability of interior points on the storage-bandwidth tradeoff", IEEE Transactions on Information Theory, 2011, 58(3): 1837-1852.